\documentclass[a4paper]{article}
\usepackage[a4paper]{geometry}
\usepackage{amsfonts}
\usepackage{latexsym}
\usepackage{amsmath}
\usepackage{amssymb}
\usepackage{mathbbol}
\usepackage{graphicx}
\author{Friedrich Hubalek\thanks{Vienna University of Technology, Austria}
\and Petra Posedel\thanks{Department of Mathematics, Faculty of Economics \& Business Zagreb, Croatia}
}
\title{Asymptotic analysis for a simple explicit estimator in
Barndorff-Nielsen and Shephard stochastic volatility models\thanks{
We thank Ole Barndorff-Nielsen, Michael S\o rensen, Bent Nielsen and S\o ren Johansen
for helpful comments. We thank Mathieu Kessler for making his
PhD-thesis available to us.
Financial support from the Austrian Science Fund
(FWF) under grant P15889 is gratefully acknowledged.}
}
\def\proof{{\textbf Proof:~}}
\def\qed{\mbox{}\hfill$\Box$}
\date{\today}
\def\Levy{L\'evy}

\def\binom#1#2{{{#1}\choose{#2}}}
\def\asto{\stackrel{a.s.}{\longrightarrow}}
\def\inlawto{\stackrel{\mathcal D}{\longrightarrow}}

\def\Var{\mbox{Var}}
\newtheorem{theorem}{Theorem}
\newtheorem{remark}{Remark}
\newtheorem{lemma}{Lemma}

\newtheorem{proposition}{Proposition}

\newtheorem{assumption}{Assumption}

\newcommand{\dis}{\displaystyle}
\newcommand{\lijepof}{\mathcal{F}}

\newcommand{\dz}{\\ \parindent 0pt \emph{\textbf{Proof}:}
\parindent 6mm}
\begin{document}
\maketitle
\begin{abstract}
We provide a simple explicit estimator for discretely observed
Barndorff-Nielsen and Shephard models, prove rigorously
consistency and asymptotic normality based on the single assumption
that  all moments of the stationary distribution of the variance
process are finite, and give explicit expressions for the asymptotic
covariance matrix.

We develop in detail the martingale estimating function approach
for a bivariate model, that is not a diffusion, but admits jumps.
We do not use ergodicity arguments.

We assume that both, logarithmic returns and instantaneous
variance are observed on a discrete grid of fixed width, and the
observation horizon tends to infinity. As the instantaneous
variance is not observable in practice, our results cannot be
applied immediately. Our purpose is to provide a theoretical
analysis as a starting point and benchmark for further
developments concerning optimal martingale estimating functions,
and for theoretical and empirical investigations, that replace the
variance process with a substitute, such as number or volume of
trades or implied variance from option data.
\end{abstract}
\subsubsection*{KEYWORDS:} Martingale estimating functions,
stochastic volatility models with jumps,
consistency and asymptotic normality
\section{Introduction}
In \cite{BNS} Barndorff-Nielsen and Shephard introduced a class of
stochastic volatility models in continuous time, where the
instantaneous variance follows an Ornstein-Uhlenbeck type process
driven by an increasing \Levy{} process. Those models allow
flexible modelling, capture many stylized facts of financial time
series, and yet are of great analytical tractability. For further
information see also \cite{BNNS}. BNS-models, as we will call them
from now on, are affine models in the sense of \cite{DPS} and
\cite{DFS}, where the associated Riccati type equations can be
solved up to quadrature in general. In several concrete cases the
integration can be performed explicitly in closed form in terms of
elementary functions, see \cite{NV} and \cite{VT}.

BNS-models have been studied from various points of view in
mathematical finance and related fields. In \cite{NV} option
pricing and structure preserving martingale measures are studied.
In \cite{BK2005,BMB2005,BG2005,RS2006} the minimal entropy
martingale measure is investigated. The papers
\cite{BKR2003,Lin2006} address the portfolio optimization problem.
Bayesian/MCMC/computer intensive estimation is already in the
seminal paper \cite{BNS}, and in the works
\cite{RPD2004,GS2001,FSS2001,tHo2003}. The papers
\cite{Jam2005,Jam2006} exploit the analytical tractability to
develop maximum likelihood estimation using the results of
\cite{cif1,cif2} for Dirichlet processes. BNS models are also
treated in the textbooks \cite{CT,Sch}.

Unfortunately, it seems that statistical estimation of the model
is the most difficult problem, and most of the work in that area
is focused on computationally intensive methods.

The contributions of the present paper are as follows: first we
develop a simple and explicit estimator for BNS models. Secondly,
we give rigorous proofs of its consistency and asymptotic
normality. In doing so we compute explicitly the asymptotic
covariance matrix and develop to that purpose formulas for
arbitrary bivariate integer moments of returns and variance.
Thirdly we provide a detailed application of the theory of
martingale estimating functions in a non-diffusion setting,
including numerical illustrations.

The literature on estimation for discretely observed diffusions is
vast, a few references are
\cite{di1,di2,di3,di4,di6,di5,di7,di8,di9,di10,di11,di12,di13}. In
particular, the martingale estimating function approach is used,
developed and studied for example in \cite{SO99}, \cite{HT},
\cite{Soerensen-Berlin}. In the diffusion setting the major
difficulty is that the transition probabilities are not known and
are difficult to compute. In contrast to that, the characteristic
function of the transition probability is known in closed form for
many BNS models and the transition probability can be computed
with Fourier methods with high precision. Yet the model exhibits
other peculiarities, see the remarks in section~\ref{remarks}.

In the present paper we explore the joint distribution of
logarithmic returns~$X$ and the instantaneous variance~$V$
supposing that both processes can be observed in discrete time.
Since the joint conditional moment-generating function of $(X,V)$
is known in closed form we obtain closed form expressions for the
joint conditional moments up to any desired order which yields a
sequence of martingale differences. We employ then the large
sample properties, in particular the strong law of large numbers
for martingales and martingale central limit theorem. In this way
we do not  need ergodicity, mixing conditions, etc.\footnote{Let
us mention though, that the martingale strong law and the ergodic
theorem have similar proofs and can be derived from a common
source, \cite{Rao73}.}

The remainder of the paper is organized as follows: in
section~\ref{continuous} we describe the class of BNS models in
continuous time and present two concrete examples, the $\Gamma-$OU
and IG-OU model. In section~\ref{discrete} we introduce the
quantities observed in discrete time that are used for estimation.
Section~\ref{remarks} contains some remarks of particular features
of the model and its estimation. In section~\ref{main} we present
the estimating equations, their explicit solution which is our
estimator and prove its consistency and asymptotic normality. 
In section~\ref{num} we present numerical illustrations. In
section~\ref{further} we sketch further and alternative
developments, in particular considering the issue that volatility
is typically not observed in discrete time. Explicit moment
calculations of any order can be found in
Appendix~\ref{explicit}. 
In Appendix~\ref{multivariate} we provide for the reader's
convenience a simple multivariate martingale central limit
theorem.
\section{The model}
\subsection{The continuous time model\label{continuous}}
\subsubsection{The general setting}
As in Barndorff-Nielsen and Shepard \cite{BNS}, we assume that the
price process of an asset $S$ is defined on some filtered
probability space $\left(\Omega,\mathcal F,(\mathcal
F_t)_{t\geq0},P\right)$ and is given by $S_t=S_0\exp(X_t)$ with
$S_0>0$ a constant. The process of logarithmic returns~$X$ and the
instantaneous variance process~$V$ satisfy
\begin{equation}
dX(t)=(\mu+\beta V(t-))dt+\sqrt{V(t-)}dW_\theta(t)+\rho
dZ_\lambda(t),\quad X(0)=0.
\end{equation}
and
\begin{equation}\label{dV}
dV(t)=-\lambda V(t-)dt+dZ_\lambda(t),\quad V(0)=V_0,
\end{equation}
where the parameters $\mu, \beta, \rho$ and $\lambda$ are real
constants with $\lambda>0.$ The process $W$ is a standard Brownian
motion, the process $Z$  is an increasing L\'{e}vy
process, and we define $Z_\lambda(t)=Z(\lambda t)$ for notational simplicity.
Adopting the terminology introduced by Barndorff-Nielsen and
Shepard, we will refer to $Z$ as the \emph{background driving
L\'{e}vy process} (BDLP).
The Brownian motion $W$ and the BDLP $Z$
are independent and $(\mathcal F_t)$ is assumed to be the usual
augmentation of the filtration generated by the pair $(W,Z_\lambda)$.
The random variable $V_0$ has a self-decomposable distribution
corresponding to the BDLP such that the process $V$ is strictly
stationary and
\begin{equation}
E[V_0]=\zeta,\qquad \Var[V_0]=\eta.
\end{equation}
To shorten the notation we introduce the parameter vector
\begin{equation}
\theta=(\lambda,\zeta,\eta,\mu,\beta,\rho)^\top,
\end{equation}
and the bivariate process
\begin{equation}
\mathbf X=(X,V).
\end{equation}
If the distribution of $V_0$ is from a particular class $D$  then
$\mathbf{X}$ is called a BNS-DOU($\theta$) model.

The process $(X_t,V_t)_{t\geq0}$ is clearly Markovian.
\subsubsection{The $\Gamma$-OU model\label{Sec-GaOU}}
The $\Gamma$-OU model is obtained by constructing the BNS-model with stationary
gamma distribution, $V_0\sim\Gamma(\nu,\alpha)$, where the parameters
are $\nu>0$ and $\alpha>0$.
The corresponding background driving \Levy{} process~$Z$ is
a compound Poisson processes with intensity $\nu$ and jumps
from the exponential distribution with parameter $\alpha$.
Consequently both processes $Z$ and $V$
have a finite number of jumps in any finite time interval.

For the $\Gamma$-OU model it is more convenient to work with the
parameters $\nu$ and $\alpha$. The connection to the generic
parameters used in our general development is given by
\begin{equation}\label{al-nu}
\zeta=\frac{\nu}{\alpha},
\qquad
\eta=\frac{\nu}{\alpha^2}.
\end{equation}
As the gamma distribution admits exponential moments we have
integer moments of all orders and our Assumption~\ref{moments}
below is satisfied.
\subsubsection{The IG-OU model}
The IG-OU model is obtained by constructing the BNS-model with
stationary inverse Gaussian distribution,
$V_0\sim(\delta,\gamma)$, with parameters $\delta>0$ and
$\gamma>0$.

The corresponding background driving \Levy{} process is the
sum of an IG($\delta/2,\gamma)$ process and an independent
compound Poisson process with intensity $\delta\gamma/2$ and jumps
from an $\Gamma(1/2,\gamma^2/2)$ distribution.
Consequently both processes $Z$ and $V$
have infinitely many jumps in any finite time interval.

For the IG-OU model it is more convenient to work with the
parameters $\delta$ and $\gamma$. The connection to the generic
parameters used in our general development is given by
\begin{equation}\label{de-ga}
\zeta=\frac{\delta}{\gamma},
\qquad
\eta=\frac{\delta}{\gamma^3}.
\end{equation}
As the inverse Gaussian distribution admits exponential moments we have
integer moments of all orders and our Assumption~\ref{moments}
below is satisfied.
\subsection{Discrete observations\label{discrete}}
We observe returns and the variance process on a discrete grid of
points in time,
\begin{equation}
0=t_0<t_1<\ldots<t_n.
\end{equation}
This implies
\begin{equation}
V(t_i)=V(t_{i-1})e^{-\lambda(t_i-t_{i-1})}
+\int_{t_{i-1}}^{t_i}e^{-\lambda(t_i-s)}dZ_\lambda(s).
\end{equation}
Using
\begin{equation}\label{defuv}
V_i:=V(t_i),\quad
U_i:=\int_{t_{i-1}}^{t_i}e^{-\lambda(t_i-s)}dZ_\lambda(s)
\end{equation}
we have that $(U_i)_{i\geq1}$ is a sequence of independent random
variables, and it is independent of~$V_0$. If the grid is
equidistant, then $(U_i)_{i\geq1}$ are iid. Observing the returns
$X$ on the grid we have
\begin{equation}
\renewcommand{\arraystretch}{2}
\begin{array}{l}
X(t_i)-X(t_{i-1})=\mu(t_i-t_{i-1})+\beta(Y(t_i)-Y(t_{i-1}))
\\
\mbox{\hspace{3cm}}\displaystyle{}+\int_{t_{i-1}}^{t_i}\sqrt{V(s-)}dW(s)
+\rho(Z_\lambda(t_i)-Z_\lambda(t_{i-1})),
\end{array}
\end{equation}
where
$$ Y(t)=\int_0^t V(s-)ds$$ is the integrated variance process.
This suggests introducing the discrete time quantities
\begin{equation}\label{defx}
X_i=X(t_i)-X(t_{i-1}),\quad Y_i=Y(t_i)-Y(t_{i-1}),\quad
Z_i=Z_\lambda(t_i)-Z_\lambda(t_{i-1})
\end{equation}
and
\begin{equation}
W_i=\frac1{\sqrt{Y_i}}\int_{t_{i-1}}^{t_i}\sqrt{V(s-)}dW(s).
\end{equation}
Furthermore, it is also convenient to introduce the discrete
quantity
\begin{equation}\label{defs}S_i=\frac{1}{\lambda}(Z_i-U_i).
\end{equation}
It is not difficult to
see (conditioning!) that $(W_i)_{i\geq1}$ is an iid $N(0,1)$
sequence independent from all other discrete quantities. We note
also that $(U_i,Z_i)_{i\geq1}$ is a bivariate iid sequence, but
$U_i$ and $Z_i$ are obviously dependent.

From now on, for notational simplicity, we consider the
equidistant grid with
\begin{equation}t_k=k\Delta,
\end{equation}
where $\Delta>0$ is fixed.
This implies
\begin{equation}\label{AR}V_i=\gamma V_{i-1}+U_i
\end{equation} and
\begin{equation}\label{defy}Y_i=\epsilon V_{i-1}+S_i,
\end{equation}where
\begin{equation}
\gamma=e^{-\lambda\Delta},\qquad\epsilon=\frac{1-\gamma}{\lambda}.
\end{equation}
Furthermore,
\begin{equation}\label{defxd}
X_i=\mu\Delta+\beta Y_i+\sqrt{Y_i}W_i+\rho Z_i.
\end{equation}
The sequence $(X_i,V_i)_{i\geq0}$ is clearly Markovian.
From now on we assume all moments of the stationary distribution
of $V_0$ exist.
\begin{assumption}\label{moments}
\begin{equation}
E[V_0^n]<\infty\qquad\forall n\in\mathbb N.
\end{equation}
\end{assumption}
In the estimating context we assume all moments are
finite with respect to all probability measures $P_\theta, \theta\in\Theta$ under
consideration, where~$\Theta$ is the parameter space.

No other assumptions are made, and all conditions required for
consistency and asymptotic normality of our estimator will be
proven rigorously from that assumption.
\begin{proposition}
We have for all $n\in\mathbb N$ that
\begin{equation}
E[Z_1^n]<\infty,\quad E[U_1^n]<\infty,\quad E[S_1^n]<\infty,
\end{equation}
and
\begin{equation}
E[Y_1^n]<\infty,\quad E[W_1^n]<\infty,\quad E[X_1^n]<\infty.
\end{equation}
Consequently the expectation of any (multivariate) polynomial in
$Z_1,U_1,S_1,\sqrt{Y_1},W_1,X_1$ exists under $P_\theta$.
\end{proposition}
\proof We will use repeatedly the well-known relation between the
existence of moments and the differentiability of the
characteristic function of a random variable, see
\cite[Theorem~8.4.1, p.295f]{CT1997}, for example.

Let $\phi(t)$ denote the characteristic function of $V_0$. By
assumption $E_\theta[V_0^n]<\infty$ for all $n\in\mathbb N$. Thus
$\phi(t)$ is arbitrarily many times differentiable. The law of
$V_0$ is self-decomposable, thus infinitely divisible and
$\phi(t)\neq0$ for all $t\in\mathbb R$. Thus the Fourier cumulant
function $\kappa(t)=\log \phi(t)$ is arbitrarily many times
differentiable. It follows from \cite[equation (12)]{BNS}, that
the characteristic function of $Z(1)$ is
$\psi(t)=\exp(t\kappa'(t))$. Thus $\psi(t)$ is arbitrarily many
times differentiable and consequently $E[Z(1)^n]<\infty$, for all
$n\in\mathbb N$. As $Z$ is a \Levy{} process this implies
$E[Z(\lambda)^n]<\infty$, and as $Z_1=Z(\lambda)$ we have shown
$E[Z_1^n]<\infty$, for all $n\in\mathbb N$.

From (\ref{defuv}) and (\ref{defs}) we have $U_1\leq Z_1$ and
$S_1\leq \lambda^{-1}Z_1$ so $E[U_1^n]<\infty$ and
$E[S_1^n]<\infty$ for all $n\in\mathbb N$. As $W_1$ has a standard
normal distribution it follows trivially $E[W_1^n]<\infty$ for all
$n\in\mathbb N$. Repeated application of the binomial resp.\
multinomial theorem, the H\"older and the Cauchy-Schwarz
inequalities yields $E[Y_1^n]<\infty$ and $E[X_1^n]<\infty$ for
all $n\in\mathbb N$, and the final conclusion for
polynomials.~\qed

Let us remark that, by the
stationarity, the above result holds also for
$Z_i,U_i,S_i,\sqrt{Y_i},W_i,X_i$ instead of
$Z_1,U_1,S_1,\sqrt{Y_1},W_1,X_1$, where $i\in\mathbb N$ is
arbitrary.
\subsection{Some remarks\label{remarks}}
Most work on estimating functions is developed for diffusions, see
for example
\cite{Soerensen-Berlin,di1,di2,di3,di4,di6,di5,di7,di8,di9,di10,di11,di12,di13},
although it is often remarked that the results extend to Markov
chains. Yet the models under consideration here display several
peculiarities.

One assumption that is usually made is that the transition
probabilities under $P_\theta$ have the same support for each
$\theta.$ Typically the support of the conditional distribution of
$V_1$ in a BNS model given $V_0=v$ is
$(ve^{-\lambda\Delta},+\infty)$ under $P_\theta$, thus depends on
$\theta$. This does not affect our analysis. The experiment is not
homogeneous, cf.\cite{Strasser}.

If the BDLP is a compound Poisson process, as in the $\Gamma-$OU
case, we have the atom of the conditional distribution of $V_1$
given $V_0=v$ under $P_\theta$ at the parameter dependent position
$ve^{-\lambda\Delta}$. Consequently no dominating measure exists and
maximum likelihood cannot be {\em defined} in the usual way. There
is an alternative definition covering that case, cf.~\cite{KW1956,Joh1978},
but we have not exploited that direction
further. See also \cite{NiSh}. This problem does not appear with
an infinite activity BDLP such as in the IG-OU model and standard
maximum likelihood estimation could be studied.

The description given in sections~\ref{continuous}
and~\ref{discrete} provides a BNS model for each~$\theta$, but not
a statistical experiment as it is taken as a starting point in
section~\ref{main}. The reason is that the processes~$X$ and~$V$
will depend on~$\theta$. This can be avoided by introducing a
\emph{statistical experiment generated by a BNS model}. In analogy
to the statistical experiment generated by a diffusion, see
\cite{SS2000}. This means we take the distribution of~$X$ and~$V$
on the Skorohod space
$\big(\mathbb{D}^2,\mathcal{B}(\mathbb{D}^2)\big)$ under each
$P_\theta$ as a starting point.
\section{The simple explicit estimator\label{main}}
\subsection{The simple estimating equations and their explicit solution}
For estimation purposes we consider a probability space on which a
parameterized family of probability measures is given:
\begin{equation}
\big(\Omega,\lijepof,\big\{P_\theta:\theta\in\Theta\big\}\big),
\end{equation}
where $\Theta=\{\theta\in\mathbb R^6:\theta^1>0,\theta^2>0,\theta^3>0\}$.
The data is generated under the true probability measure $P_{\theta_0}$ with some $\theta_0\in\Theta$.
The expectation with respect to $P_\theta$  is denoted  by $E_\theta[.]$ and with
respect to $P_{\theta_0}$ simply by $E[.]$.

We assume there is a process~$\mathbf{X}$ that is BNS-DOU($\theta$) under~$P_\theta$.
We want to find an estimator for~$\theta_0$ using observations
$X_1,\ldots,X_n,V_1,\ldots,V_n$. We are interested in asymptotics
as $n\to \infty$.
To that purpose let us consider the following martingale estimating functions:
\begin{equation}\label{mef1}\renewcommand{\arraystretch}{1.5}
\begin{array}{ll}
G_n^1(\theta)=\sum_{k=1}^n\big[V_k       -f^1(V_{k-1},\theta)\big],&\qquad f^1(v,\theta)=E_\theta[V_1   |V_0=v]\\
G_n^2(\theta)=\sum_{k=1}^n\big[V_kV_{k-1}-f^2(V_{k-1},\theta)\big],&\qquad f^2(v,\theta)=E_\theta[V_1V_0|V_0=v]\\
G_n^3(\theta)=\sum_{k=1}^n\big[V_k^2     -f^3(V_{k-1},\theta)\big],&\qquad f^3(v,\theta)=E_\theta[V_1^2 |V_0=v]\\
G_n^4(\theta)=\sum_{k=1}^n\big[X_k       -f^4(V_{k-1},\theta)\big],&\qquad f^4(v,\theta)=E_\theta[X_1   |V_0=v]\\
G_n^5(\theta)=\sum_{k=1}^n\big[X_kV_{k-1}-f^5(V_{k-1},\theta)\big],&\qquad f^5(v,\theta)=E_\theta[X_1V_0|V_0=v]\\
G_n^6(\theta)=\sum_{k=1}^n\big[X_kV_k    -f^6(V_{k-1},\theta)\big],&\qquad f^6(v,\theta)=E_\theta[X_1V_1|V_0=v]
\end{array}
\end{equation}
\begin{lemma}
We have the explicit expressions
\begin{equation}
\begin{array}{l}\label{deff}
f^1(v,\theta)=\gamma v+(1-\gamma)\zeta\\
f^2(v,\theta)=\gamma v^2+(1-\gamma)\zeta v\\
f^3(v,\theta)=\gamma^2 v^2+2\gamma(1-\gamma)\zeta v+(1-\gamma)^2\zeta^2+(1-\gamma^2)\eta\\
f^4(v,\theta)=\beta\epsilon v+\mu\Delta+\beta(1-\epsilon)\zeta+\rho\lambda\zeta\\
f^5(v,\theta)=\beta\epsilon v^2+(\mu\Delta+\beta(1-\epsilon)\zeta+\rho\lambda\zeta)v\\
f^6(v,\theta)=\beta\epsilon\gamma v^2+((\mu\Delta+\beta(1-\epsilon)\zeta+\rho\lambda\zeta)\gamma+\beta\epsilon(1-\gamma)\zeta)v+(1-\epsilon)(1-\gamma)\zeta^2+\epsilon^2\lambda\eta
\end{array}
\end{equation}
\end{lemma}
\proof  The formulas are special cases of the general moment
calculations given in Appendix~\ref{explicit}. In order to
demonstrate the basic idea we will prove the statements for two
special and simple cases here, namely for $f^1(v,\theta)$ and
$f^4(v,\theta)$. From (\ref{AR}) it follows that
\begin{equation}E_\theta[V_1|V_0=v]=\gamma v+E_\theta[U_1]
\end{equation}
and from the stationarity of $V$ we have
\begin{equation}E_\theta[U_1]=(1-\gamma)E_\theta(V_0)=(1-\gamma)\zeta.
\end{equation}
Furthermore, from (\ref{defxd}) and the fact that $E[W_1]=0,$ it
follows that
\begin{equation}\label{f4}E_\theta[X_1|V_0=v]=\mu\Delta+\beta E_\theta[Y_1|V_0=v]+\rho E_\theta[Z_1|V_0=v].
\end{equation}
But, from (\ref{defy}) we have that
\begin{equation}E_\theta[Y_1|V_0=v]=\epsilon
v+\frac{1}{\lambda}E_\theta[Z_1-U_1]=\epsilon v+\zeta(1-\epsilon),
\end{equation}and
\begin{equation}E_\theta[Z_1]=\lambda\zeta.
\end{equation}
So, from (\ref{f4}) it follows that
\begin{equation}E_\theta[X_1|V_0=v]=\mu\Delta+\beta\epsilon
v+\beta(1-\epsilon)\zeta+\rho\lambda\zeta.
\end{equation}
\qed

The estimator $\hat{\theta}_n$ is obtained by solving the
estimating equation $G_n(\theta)=0$ and it turns out that this
equation has a simple explicit solution.
\begin{proposition}\label{ESTIMATOR}
The estimating equation $G_n(\hat\theta_n)=0$ admits for every $n\geq2$
on the event
\begin{equation}
C_n=\big\{\xi_n^2-\xi_n^1\upsilon_n^1>0,\upsilon_n^2-(\upsilon_n^1)^2>0\big\}
\end{equation}
a unique solution $\hat\theta_n=(\lambda_n,\zeta_n,\eta_n,\beta_n,\rho_n,\mu_n)$
that is given by
\begin{equation}\label{thn}
\renewcommand{\arraystretch}{1.5}
\begin{array}{l}
\gamma_n=(\xi_n^2-\xi_n^1\upsilon_n^1)/(\upsilon_n^2-(\upsilon_n^1)^2);
\\
\zeta_n=(\xi_n^1-\gamma_n\upsilon_n^1)/(1-\gamma_n);
\\
\eta_n =((\xi_n^3-(\xi_n^1)^2)-\gamma_n^2(\upsilon_n^2-(\upsilon_n^1)^2))/(1-\gamma_n^2);
\\
\lambda_n=-\log(\gamma_n)/\Delta;
\\
\epsilon_n=(1-\gamma_n)/\lambda_n;
\\
\beta_n=(\xi_n^5-\upsilon_n^1\xi_n^4)/(\epsilon_n(\upsilon_n^2-(\upsilon_n^1)^2));
\\
\rho_n=(\xi_n^6-\xi_n^4\xi_n^1-\beta_n\epsilon_n(\eta_n(1-\gamma_n)
+\gamma_n(\upsilon_n^2-(\upsilon_n^1)^2)))/(2(1-\gamma_n)\eta_n);
\\
\mu_n=(\xi_n^4-\beta_n\epsilon_n(\upsilon_n^1-\zeta_n))/\Delta-(\beta_n+\lambda_n\rho_n)\zeta_n;
\end{array}
\end{equation}
where
\begin{equation}\label{xin}
\renewcommand{\arraystretch}{2.0}
\begin{array}{lll}
\xi_n^1=\frac1n\sum\limits_{i=1}^nV_i
&
\xi_n^2=\frac1n\sum\limits_{i=1}^nV_iV_{i-1}
&
\xi_n^3=\frac1n\sum\limits_{i=1}^nV_i^2
\\
\xi_n^4=\frac1n\sum\limits_{i=1}^nX_i
&
\xi_n^5=\frac1n\sum\limits_{i=1}^nX_iV_{i-1}
&
\xi_n^6=\frac1n\sum\limits_{i=1}^nX_iV_i
\end{array}
\end{equation}
and
\begin{equation}\label{upn}
\begin{array}{ll}
\upsilon_n^1=\frac1n\sum\limits_{i=1}^nV_{i-1}
&
\upsilon_n^2=\frac1n\sum\limits_{i=1}^nV_{i-1}^2
\end{array}
\end{equation}
\end{proposition}
\proof The first three equations $G_n^j(\theta)=0$, for $j=1,2,3$
contain only the unknowns $\zeta, \eta, \lambda$ and are easily
solved. In fact we get a familiar estimator for the first two
moments and the autocorrelation coefficient of an AR(1) process.
The last three equations $G_n^j(\theta)=0$, for $j=4,5,6$ can be
seen as a linear system for the unknowns $\mu,\beta,\rho$, once
the other parameters have been determined. \qed
\begin{remark}
The exceptional set $C_n$ could be simplified to
\begin{equation}
C_n'=\big\{\xi_n^2-\xi_n^1\upsilon_n^1>0\big\}
\end{equation}
Since the jump times and the jump sizes of the BDLP are
independent, and the former have an exponential distribution it
follows that $V_0,\ldots,V_n$ is with probability one not
constant, so $P[\upsilon_n^2-(\upsilon_n^1)^2>0]=1$. But although
it can be shown that the probability of $C_n$ tends to zero, for
finite $n$ we have $P[\xi_n^2-\xi_n^1\upsilon_n^1\leq0]>0$. This
is the common phenomenon that sample moments do not share all
properties of their theoretical counterparts. For definiteness we
put~$\hat\theta_n=0$ outside~$C_n$.
\end{remark}
\subsection{Consistency}
Let us investigate the consistency of the estimator from the
previous section. First, we will need the following lemma.
\begin{lemma}\label{ogr}
For every $k\geq 1$ and $p>0$
\begin{equation}
\frac{V_n^k}{n^p}\asto0\qquad as \quad
n\to \infty.
\end{equation}
\end{lemma}
\proof  The random variables $\dis \big\{V_n,\ n\geq
1\big\}$ are identically distributed and $\dis m_k=E[\vert
V_1^k\vert]<\infty$ for all $\dis k\geq1.$ Thus we are in the
situation of~\cite[Exercise~2.1.2(i), p.14]{Stout}.

Let $k\geq 1$ and $\epsilon>0$ be arbitrarily chosen. Taking any
integer  $\alpha>1/p$ and using the Chebyshev inequality we obtain
\begin{equation}\sum_{n=1}^\infty
P\bigg(\bigg\vert\frac{V_n^k}{n^p}\bigg\vert>\epsilon\bigg)\leq\sum_{n=1}^\infty\frac{E\big\vert
V_n^k\vert^\alpha}{n^{\alpha p}\epsilon^\alpha}
\leq\sum_{n=1}^\infty\frac{m_{k\alpha}}{n^{\alpha
p}\epsilon^{\alpha}}<\infty.
\end{equation}
Therefore from the Borel-Cantelli lemma it follows that
$P\big(\limsup_n{n^{-p}}{|V_n^k|}>\epsilon\big)=0$.
\qed
\begin{lemma}\label{lemv}
We have for all $k\in\mathbb N$ that
\begin{equation}
\frac1n\sum_{i=1}^nV_i^k\asto E[V_1^k],
\end{equation}
as $n\to\infty$.
\end{lemma}
\proof We will prove this statement by
induction. \\
\noindent $\dis (1)\ k=1.$ Let us define
$$H_i=V_i-E(V_i\vert V_{i-1}),\qquad i\geq 1.$$
Obviously, $(H_i,\ i\geq 1)$ is a sequence of martingale
differences and is therefore uncorrelated. Using expressions
(\ref{defuv}) and (\ref{AR}) we obtain
\begin{eqnarray*}E\big[H_i^2\big]&=&E(V_i^2)-E\big[E(V_i|V_{i-1})^2\big]\\
&=&(1-\gamma^2)E(V_0^2)-2\gamma E(U_1)E(V_0)+E(U_1^2),
\end{eqnarray*}so  $\dis E\big[H_i^2\big]$ have a common bound
for every $i\geq 1.$ Since the assumptions of the Theorem 5.1.2
from \cite{Chung.2001} are satisfied, it follows that
\begin{equation*}\frac{1}{n}\sum_{i=1}^nV_i-\frac{1}{n}\sum_{i=1}^nE(V_i|V_{i-1})\asto0,\qquad
as\quad n\to\infty.
\end{equation*}But using again definition (\ref{AR}), the last
expression is equivalent to
\begin{equation*}\frac{1-\gamma}{n}\sum_{i=1}^nV_i+\frac{\gamma(V_n-V_0)}{n}-E(U_1)\asto0,\qquad
as\quad n\to\infty.
\end{equation*}Finally, using the result of the previous lemma and the fact that $E[U_1]=(1-\gamma)E[V_0],$ it
follows
\begin{equation*}\frac{1}{n}\sum_{i=1}^nV_i\asto E(V_0),\qquad
as\quad n\to \infty.
\end{equation*}
This completes the proof for $\dis k=1.$\\
\noindent (2)\ Suppose now that the statement of the theorem holds
for $l\leq k-1,$ i.e. $E(V_1^{k-1})<\infty$ and
\begin{equation}\label{indk1}\frac{1}{n}\sum_{i=1}^nV_i^{l}\asto E(V_0^{l}),\quad
l\leq k-1
\end{equation}when $\dis n\to \infty.$
 For $\dis k>1,$ and for $\dis i\geq 1,$ let
\begin{equation}\label{defxi}H_i^k:=V_i^k-E\big[V_i^k\vert
V_{i-1}\big]\quad\mathrm{and}\quad S_n^{k}:=\sum_{i=1}^nH_i^k.
\end{equation}Obviously, $\dis (H_i^k,\ i\geq 1)$ is a sequence of martingale
differences and in particular is uncorrelated. Moreover, due to
the strong stationarity of the volatility sequence and relations
(\ref{defxi}) and (\ref{AR}) we obtain
\begin{eqnarray}E\big[H_i^k\big]^2&=&E\big[V_i^{2k}-2V_i^kE[V_i^k\vert V_{i-1}]+E[V_i^k\vert
V_{i-1}]^2\big]\nonumber\\
&=&E[V_i^{2k}]-E\big[(E[V_i^k\vert V_{i-1}])^2\big]\nonumber\\
&\leq& E[V_1^{2k}]=:c_k,
\end{eqnarray}$\dis c_k$ denoting some constant that does not
depend on $i.$ Hence, by~\cite[Theorem~5.1.2, p.108]{Chung.2001},
it follows that $\dis \frac{S_n^{k}}{n}\asto0$ when $\dis n\to
\infty,$ which in our case, due to the definition of $\dis
S_n^{k},$ is equivalent to
\begin{equation}\label{kvrgm}\frac{1}{n}\sum_{i=1}^n
V_i^k-\frac{1}{n}\sum_{i=1}^nE\big[V_i^k\vert
V_{i-1}\big]\asto0.
\end{equation}
Using again the definition (\ref{AR}) and the independency of
$\dis U_i$ from $\dis V_{i-1},$ for $\ i\geq 1,$ we obtain
\begin{eqnarray}\frac{1}{n}\sum_{i=1}^n
V_i^k-\frac{1}{n}\sum_{i=1}^nE\big[V_i^k\vert V_{i-1}\big]
&=&\frac{1}{n}\sum_{i=1}^nV_i^k-\frac{1}{n}\sum_{i=1}^n\sum_{j=0}^k
{k\choose j}\gamma^jV_{i-1}^jE\big[U_1^{k-j}\big]\nonumber\\
&=&\frac{1-\gamma^k}{n}\sum_{i=1}^nV_i^k-\frac{\gamma^k}{n}(V_0^k-V_n^k)-\sum_{j=0}^{k-1}{k\choose
j}E\big[U_1^{k-j}\big]\frac{\gamma^j}{n}\sum_{i=1}^nV_{i-1}^j.\nonumber
\end{eqnarray}
Finally, applying the assumption of the induction, Lemma \ref{ogr}
and the statement (\ref{kvrgm}), we obtain
$$\frac{1}{n}\sum_{i=1}^nV_i^k\asto\frac{1}{1-\gamma^k}\dis\sum_{j=0}^{k-1}{k\choose
j}\gamma^jE\big[U_1^{k-j}\big]E(V_0^j)=E(V_1^k),$$ where the last
equality follows calculating $\dis E(\gamma V_0+U_1)^k$ using
(\ref{AR}).
\qed

In the next lemma we extend the strong law of large numbers for $(V_i^p,i\geq1)$
to more general sequences.
\begin{lemma}\label{lemvx}
For all integers $p,q,r\geq0$ we have
\begin{equation}
\frac1{n}\sum_{i=1}^nX_i^pV_i^qV_{i-1}^r\asto E\big[X_1^pV_1^qV_0^r\big]
\end{equation}
as $n\to \infty$.
\end{lemma}
\proof: Let
$$M_i=X_i^pV_i^qV_{i-1}^r-E\big[X_i^pV_i^qV_{i-1}^r\vert V_{i-1}\big].$$
Obviously, $(M_i,i\geq 1)$ is
a sequence of martingale differences, and in particular
it is uncorrelated.
It is stationary and $E[M_1^2]<\infty$.
So we can use again \cite[Theorem~5.12]{Chung.2001} to show
$$
\frac1{n}\sum_{i=1}^nX_i^pV_i^qV_{i-1}^r-\frac{1}{n}\sum_{i=1}^nE\big[X_i^pV_i^qV_{i-1}^r|V_{i-1}\big]\asto0.
$$
The conditional expectation $E\big[X_i^pV_i^qV_{i-1}^r|V_{i-1}\big]$ is a polynomial in $V_{i-1}$, namely
$$
E\big[X_i^pV_i^qV_{i-1}^r|V_{i-1}\big]=
\sum_{k=0}^{p+q}\phi_{pqk}V_{i-1}^{k+r}.
$$
This is shown in section A.5 in Appendix~\ref{explicit} where the
coefficients $\phi_{pqk}$ are explicitly calculated.  Applying
Lemma~\ref{lemv} yields
$$
\frac1n\sum_{i=1}^nE\big[X_i^pV_i^qV_{i-1}^r|V_{i-1}\big]\asto
\sum_{k=0}^{p+q}\phi_{pqk}E[V_0^{k+r}].
$$
As we have
\begin{equation}
E[X_1^pV_1^qV_0^r]=E[E[X_1^pV_1^qV_0^r\vert V_0]]=
\sum_{k=0}^{p+q}\phi_{pqk}E[V_0^{k+r}],
\end{equation}
the proof is completed.~\qed
\begin{theorem} We have $P(C_n)\to 1$ when $n\to\infty$ and the
estimator $\hat{\theta}_n$ is consistent on $\dis C_n$, namely
$$\hat{\theta}_n\asto\theta_0$$ on $C_n$
as\ $n\to \infty.$
\end{theorem}
\proof  Using the results of Lemma \ref{lemv} it easily follows
that
\begin{equation}
\xi_n^2-\xi_n^1\upsilon_n^1\to Cov(V_1,V_0)>0,
\end{equation}
so~$P(C_n)\to1$ as $n\to\infty$.

Using again the results of Lemma~\ref{lemv} it follows that the
empirical moments in~(\ref{xin}) and~(\ref{upn}) converge to their
theoretical counterparts, $\xi_n^i\asto\xi^i$ and
$\upsilon_n^i\asto\upsilon^i$, where
\begin{equation}
\begin{array}{l}
\xi^1=\zeta,\\
\xi^2=\zeta^2+\gamma\eta,\\
\xi^3=\zeta^2+\eta,\\
\xi^4=\mu+(\beta+\lambda\rho)\zeta,\\
\xi^5=\mu\zeta+(\beta+\lambda\rho)\zeta^2+\beta\epsilon\eta,\\
\xi^6=\mu\zeta+(\beta+\lambda\rho)\zeta^2+(\beta+2\rho\lambda)\epsilon\eta,
\end{array}
\qquad
\begin{array}{l}
\upsilon^1=\zeta,\\
\upsilon^2=\zeta^2+\eta.
\end{array}
\end{equation}
Plugging the limits into~(\ref{thn})
shows, after a short mechanical calculation, that the estimator
is in fact consistent. \qed
\subsection{Asymptotic normality}
For a concise vector notation we introduce
\begin{equation}
\Xi_k=(V_k,V_kV_{k-1},V_k^2,X_k,X_kV_{k-1},X_kV_k)^\top,
\end{equation}
and write the estimating equations in the form
\begin{equation}
G_n^i(\theta)=\sum_{k=1}^n\big[\Xi_k^i-f^i(V_{k-1},\theta)\big],
\qquad i=1,\ldots,6
\end{equation}
and $f^i(v,\theta)$ given by (\ref{deff}). We write
\begin{equation}
f^i(v,\theta)=\sum_{\ell=r_i}^{p_i+r_i+q_i}\phi^i_{\ell}(\theta)v^{\ell}.
\end{equation}
with
\begin{equation}
p=(0,0,0,1,1,1),\quad q=(1,1,1,0,0,1),\quad r=(0,1,0,0,1,0)
\end{equation}
and
\begin{equation}
\begin{array}{lll}
\phi^1_1(\theta)=\gamma              & \phi^1_0(\theta)=(1-\gamma)\zeta\\
\phi^2_2(\theta)=\gamma              & \phi^2_1(\theta)=(1-\gamma)\zeta \\
\phi^3_2(\theta)=\gamma^2            & \phi^3_1(\theta)=2\gamma(1-\gamma)\zeta                                                             & \phi^3_0(\theta)=(1-\gamma)^2\zeta^2+(1-\gamma^2)\eta\\
\phi^4_1(\theta)=\beta\epsilon       & \phi^4_0(\theta)=\mu+\beta(1-\epsilon)\zeta+\rho\lambda\zeta\\
\phi^5_2(\theta)=\beta\epsilon       & \phi^5_1(\theta)=\mu+\beta(1-\epsilon)\zeta+\rho\lambda\zeta\\
\phi^6_2(\theta)=\beta\epsilon\gamma &
\phi^6_1(\theta)=((\mu+\beta(1-\epsilon)\zeta+\rho\lambda\zeta)\gamma+\beta\epsilon(1-\gamma)\zeta)
&
\phi^6_0(\theta)=(1-\epsilon)(1-\gamma)\zeta^2+\epsilon^2\lambda\eta
\end{array}
\end{equation}
We will use, that $f^i(v,\theta)$ is a polynomial in $v$, and that
its coefficients $\phi$ are smooth functions in $\theta$.

We shall first prove the central limit theorem for the estimating functions.
\begin{proposition}\label{prop2CLTMEF}
We have
\begin{equation}\label{CW}
\frac{1}{\sqrt{n}}G_n(\theta_0)\inlawto N(0,\Upsilon),
\end{equation}
as $n\to\infty$, where
\begin{equation}
\Upsilon_{ij}=E\big[Cov(\Xi_1^i,\Xi_1^j|V_0)\big].
\end{equation}
\end{proposition}
\proof To show the above result, we use the multivariate
martingale central limit theorem, that is recapitulated in
Appendix~\ref{multivariate}. To that purpose we introduce the
vector martingale difference array
\begin{equation}\label{fjamd}
\chi_{n,k}=\frac{1}{\sqrt{n}}\big[\Xi_k^i-f^i(V_{k-1},\theta)\big].
\end{equation}
We have to show the two assumptions from the previous theorem.
First, we prove a multivariate Lyapuonov condition which implies
the Lindeberg condition. From (\ref{fjamd}) it follows that $\dis
\sqrt{n}\chi_{n,k}^{(j)}$ is of the form $p(V_0,V_1,X_1)$ where
$p(v_0,v_1,x_1)$ is a polynomial in $v_0,v_1,x_1$ which does not
depend on $n.$ Thus, $\dis n^2\|\chi_{n,k}\|^4$ has the same
property and from the explicit moment expression from
Appendix~\ref{explicit} it follows that
\begin{equation}
E\big[\|\chi_{n,k}\|^4|\lijepof_{k-1}\big]=\frac{1}{n^2}q(V_{k-1}),
\end{equation}
where $q(v_0)$ is a polynomial in $v_0.$ From Lemma~\ref{lemv}
it thus follows
\begin{equation}
\frac1n\sum_{k=1}^nq(V_{k-1})\asto E[q(V_0)],
\end{equation}
where the expression on the righthand side exists and is finite.
Thus the first condition of the martingale central limit theorem
is satisfied. In order to verify the second condition of the same
theorem we consider the $(i,j)-$th element of the matrix
$\chi_{n,k}\chi_{n,k}^\top$ which is given by
\begin{equation}
\frac1n\big(\Xi_k^i-f^i(V_{k-1},\theta)\big)\big(\Xi_k^j-f^j(V_{k-1},\theta)\big).
\end{equation}
This is again a polynomial in $V_{k-1},V_k$ and $X_k$ so by
Lemma~\ref{lemvx} it follows that
\begin{equation}
\frac1n\sum_{k=1}^n\big(\Xi_k^i-f^i(V_{k-1},\theta)\big)\big(\Xi_k^j-f^j(V_{k-1},\theta)\big)
\asto
E\big[\big(\Xi_k^i-f^i(V_{k-1},\theta)\big)\big(\Xi_k^j-f^j(V_{k-1},\theta)\big)\big]
\end{equation}
as $n\to\infty$.\qed
\begin{remark}A systematic method to evaluate $\Upsilon$ is given
in Appendix \ref{ApA} and the resulting explicit expressions are
given in \cite{thesis}.
\end{remark}
\begin{lemma}\label{lemaCLTM}
We have
\begin{equation}\label{CLTM}
\frac1{\sqrt{n}}\big[\xi_n-\xi\big]\inlawto N(0,\Sigma),
\end{equation}
where
\begin{equation}
\Sigma=P^{-1}\Upsilon (P^{-1})^\top
\end{equation}
and
\begin{equation}
P_{ij}=\delta_{ij}-\phi_1^i\delta_{1j}-\phi^i_2\delta_{3j}
\end{equation}
with $\delta_{ij}$ denoting the Kronecker delta.
\end{lemma}
\proof We can write
\begin{equation}
\frac1{\sqrt{n}}G_n(\theta_0)=P\sqrt{n}(\xi_n-\xi)+Q_n,
\end{equation}
with
\begin{equation}\label{Qn}
Q_n^i=\frac1{\sqrt{n}}\big[\phi_1^i(V_n-V_0)+\phi_2^i(V_n^2-V_0^2)\big].
\end{equation}
In view of Lemma \ref{ogr} above we see, that the remainder term
$Q_n$ goes to zero in probability as $n\to\infty$. As $P$ has
determinant $(1-\gamma)^2(1+\gamma)>0$ it is invertible, and we
have
\begin{equation}
\sqrt{n}(\xi_n-\xi)=P^{-1}\left(\frac1{\sqrt{n}}G_n(\theta_0)\right)+R_n
\end{equation}
with $R_n=-P^{-1}Q_n$ going to zero in probability as $n\to\infty$.
The expression $P^{-1}\left(n^{-1/2}G_n(\theta_0)\right)$
is asymptotically normal with mean $0$ and covariance matrix $\Sigma$.
An application of Slutsky's Theorem proves the lemma.
\qed

Finally, we have all the ingredients for proving the following result.
\begin{theorem}
The estimator
\begin{equation}
\hat\theta_n=(\lambda_n,\zeta_n,\eta_n,\mu_n,\beta_n,\rho_n)
\end{equation}
is asymptotically normal, namely
\begin{equation}\label{CLTtheta}
\sqrt{n}\big[\hat{\theta}_n-\theta_0\big]\inlawto N(0,T),
\end{equation}
as $n\to\infty$, where
\begin{equation}
T=D\Sigma D^{T}
\end{equation}
and $D$ is explicitly given according to (\ref{MatrixD}).
\end{theorem}
\proof  We observe from~(\ref{thn}) that
$\hat\theta_n=g(\xi_n,\upsilon_n)$, where $g$ is well defined and
continuously differentiable in a neighborhood of $(\xi,\upsilon)$.
Using the Taylor expansion in the last two variables we have
$\hat\theta_n=h(\xi_n)+S_n$, where $h$ is well defined and
continuously differentiable in a neighborhood of $\xi$, and $S_n$
goes to zero in probability in view of Lemma \ref{ogr}. Thus it
can be neglected according to Slutsky's Theorem. We apply the
delta method, see~\cite{Leh} for example, and compute the Jacobian
matrix $D$ with
\begin{equation}\label{MatrixD}
D_{ij}=\frac{\partial h_i}{\partial x_j}(\xi),\qquad i,j=1,\ldots,6.
\end{equation}
A lengthy elementary
calculation shows that the matrix has
determinant $\lambda/\big(2(1-\gamma)^2\gamma\eta^3\big)$, thus it is
invertible. \qed
\begin{remark}For comparison it is instructive to study our simple
estimator in the general framework of \cite{SO99} where the
properties of the estimator are studied without exploiting the
fact that the estimating equation allows an explicit solution.
This is done in \cite{thesis}. There the theory is extended in the
case of a bivariate Markov process and Condition~2.6 of
\cite{SO99} is proven in order to use his Corollary~2.7 and
Theorem~2.8.
\end{remark}
\section{Numerical illustrations\label{num}}
\subsection{Description of the model and its parameter values}
To illustrate the results from the previous sections numerically,
we consider the $\Gamma$-OU model from Section~\ref{Sec-GaOU},
where the variance $V$ has
a stationary gamma distribution.
We use as time unit one year consisting of 250 trading days.
The true parameters are
\begin{equation}
\nu=2.56,\quad
\alpha=64,\quad
\lambda=256,\quad
\beta=-0.5,\quad
\rho=-0.1,\quad
\mu=1.2.
\end{equation}
The parameters imply that there are on average $2.6$ jumps per day
and the jumps in the BDLP and in the volatility are exponentially
distributed with mean $0.0156$. The interpretation is, that
typically every day two or three new pieces of information arrive
and make the variance process jump. The stationary mean of the
variance is 0.04. Hence, if we define instantaneous volatility to
be the square root of the variance, it will fluctuate around 20\%
in our example. The half-life of the autocorrelation of returns is
about half a day.

\begin{figure}[h]
\begin{center}
\includegraphics[width=14cm,height=5cm]{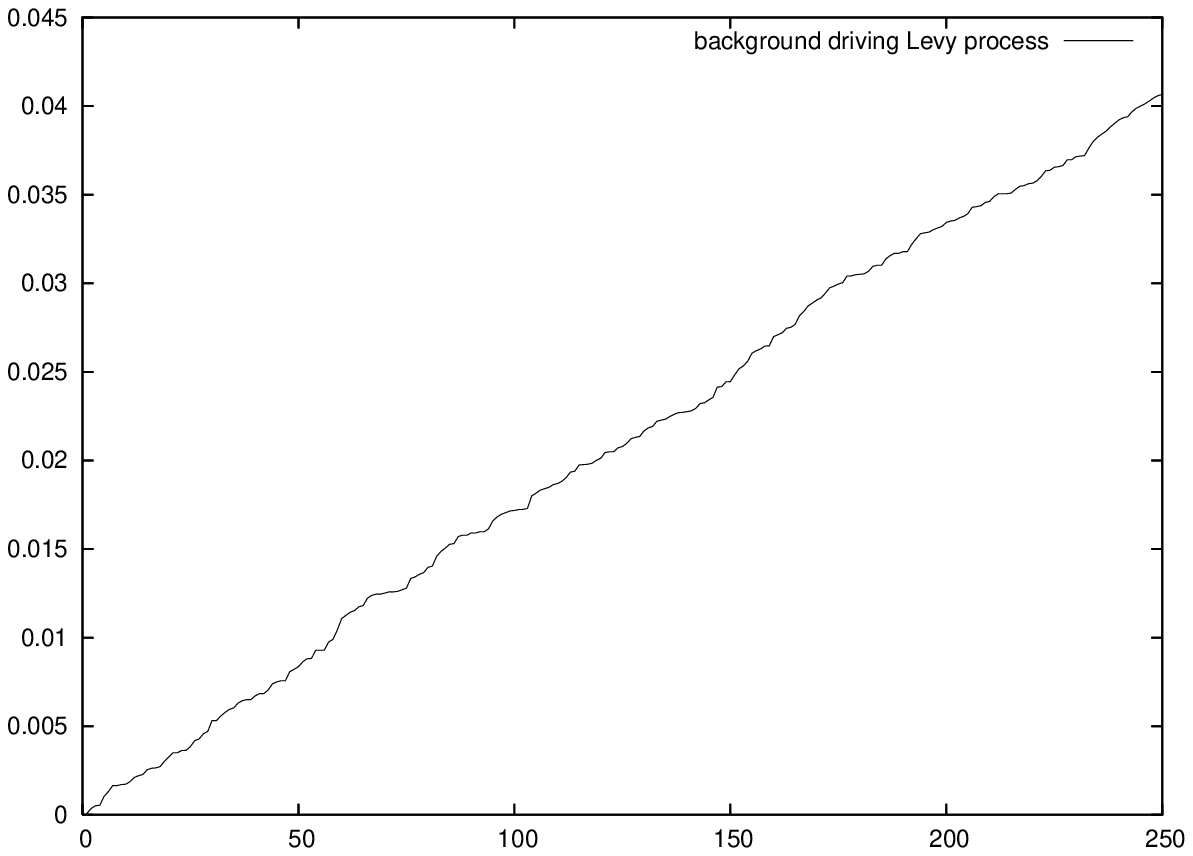}\\
\includegraphics[width=14cm,height=5cm]{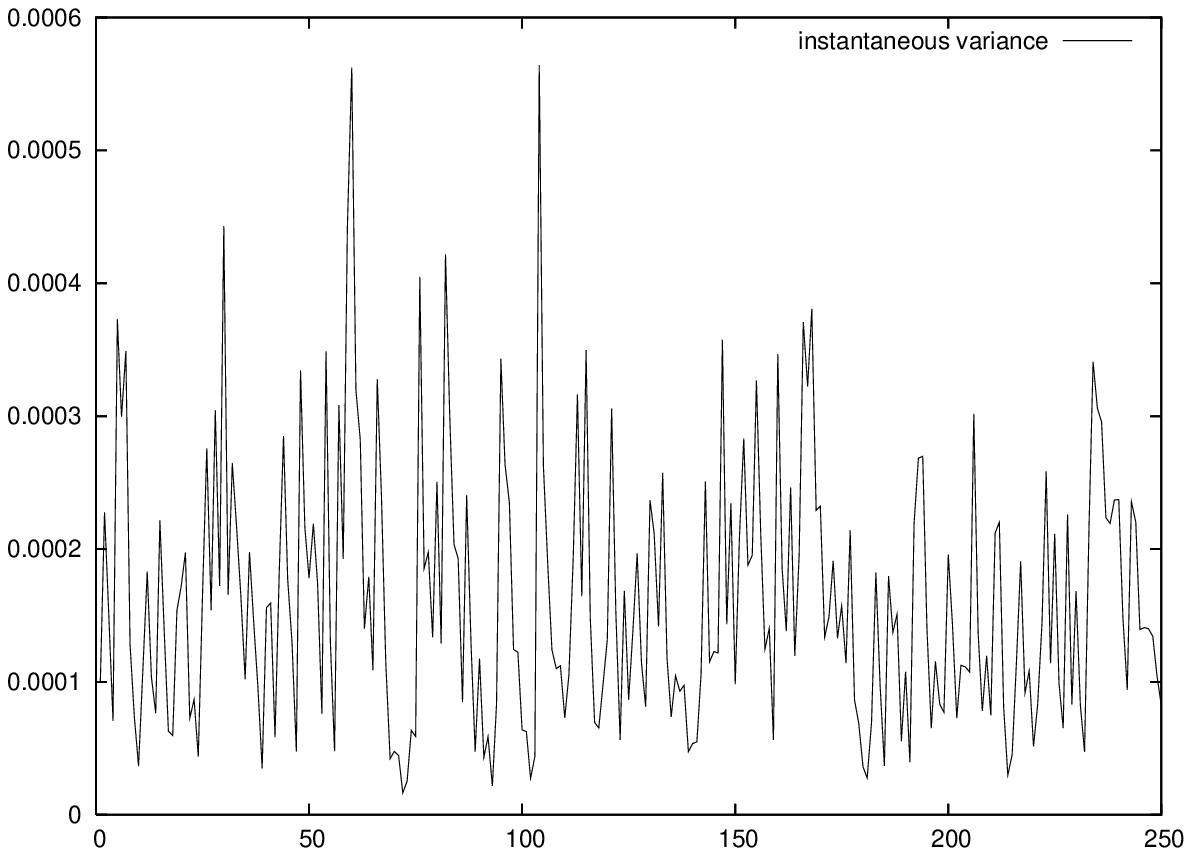}\\
\includegraphics[width=14cm,height=5cm]{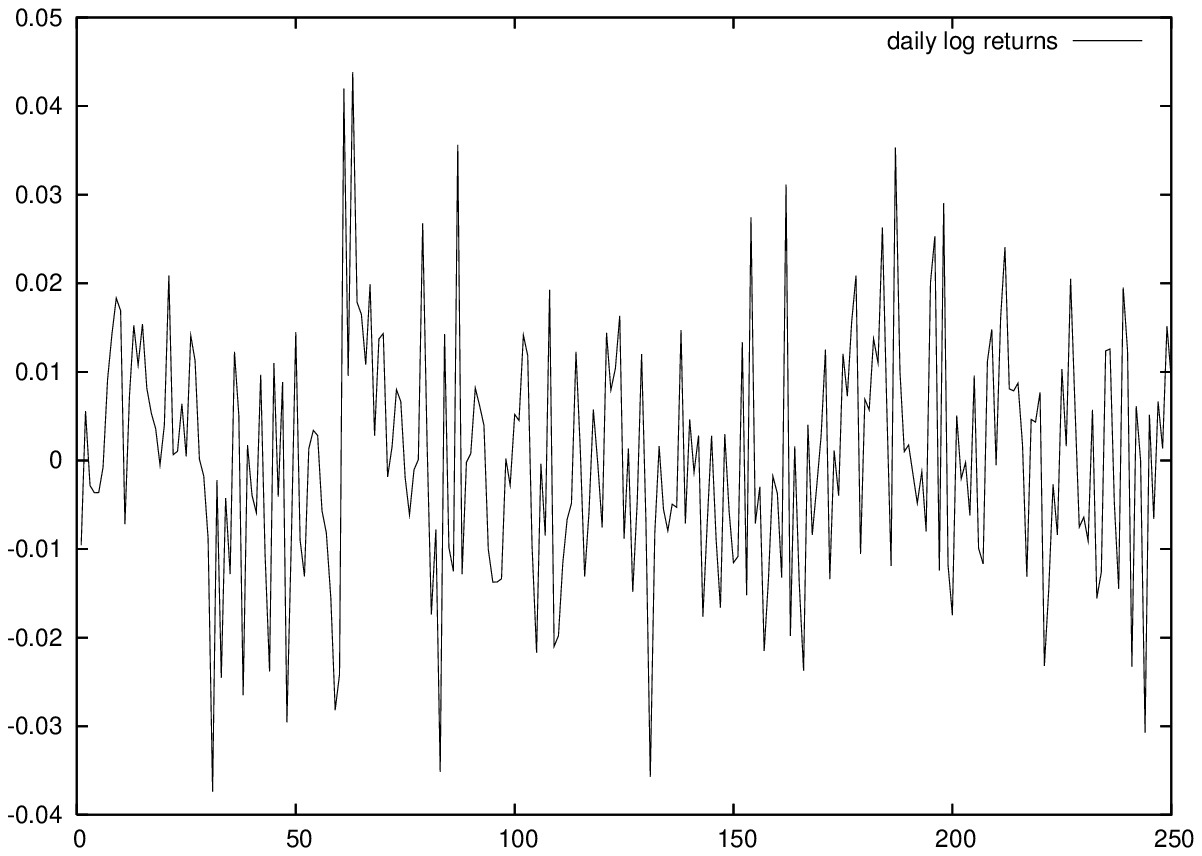}\\
\end{center}
\caption{Daily observations $Z_i,V_i,X_i$.\label{path}}
\end{figure}

In our example annual log returns have (unconditional) mean $25.6\%$ and a
annual volatility 20\%.
Figure~\ref{path} displays a simulation of one year
of daily observations from the background driving \Levy{}
process, from the instantaneous variance process, and log
returns, or more precisely, simulated realizations of $Z_i$, $V_i$, and $X_i$ for $i=1,\ldots,250$.
In~\cite{thesis} other scenarios are considered, for example,
small jumps arriving every minute, with fast decaying
autocorrelation, or few jumps per year, corresponding to
exceptional news with heavy impact on the variance process.
\subsection{The asymptotic covariance matrix of the estimator}
As our goal is an analysis of the estimator, and not an empirical
study, we do not estimate the asymptotic covariance, but
evaluate the explicit expression using the true parameters.
Denoting the vector of asymptotic standard deviations of the
estimates and the correlation matrix by $s/\sqrt{n}$
resp.~$r$ we have
\begin{equation}
s=\left[
\begin{array}{c}
4.86 \\ 125 \\ 650 \\ 7.36 \\ 253 \\ 0.526
\end{array}
\right],\qquad
r=\left[
\begin{array}{cccccc}
1 & 0.89 & 0.41 & 0.03 & 0.09 & -0.02 \\
0.89 & 1 & 0.4 & 0.03 & 0.09 & -0.03 \\
0.41 & 0.4 & 1 & 0.06 & 0.22 & 0 \\
0.03 & 0.03 & 0.06 & 1 & -0.75 & 0.06 \\
0.09 & 0.09 & 0.22 & -0.75 & 1 & -0.57 \\
-0.02 & -0.03 & 0 & 0.06 & -0.57 & 1
\end{array}
\right]
\end{equation}
\subsection{Distribution of the estimates}
\begin{figure}[h]\label{his}
$$
\begin{array}{cc}
\includegraphics{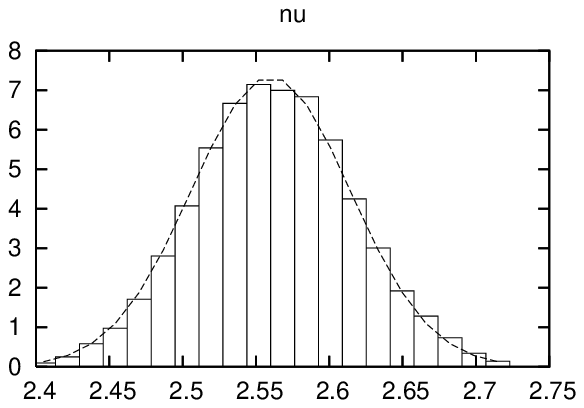} & \includegraphics{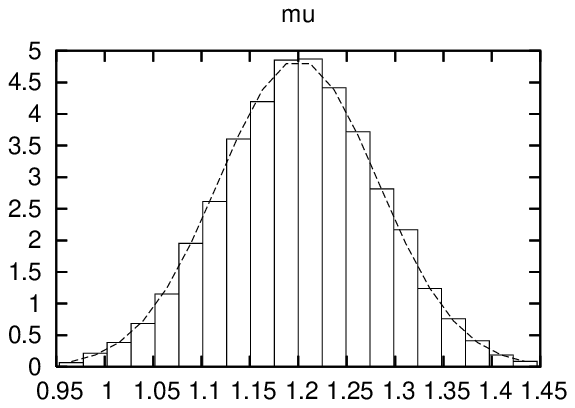}\\
\includegraphics{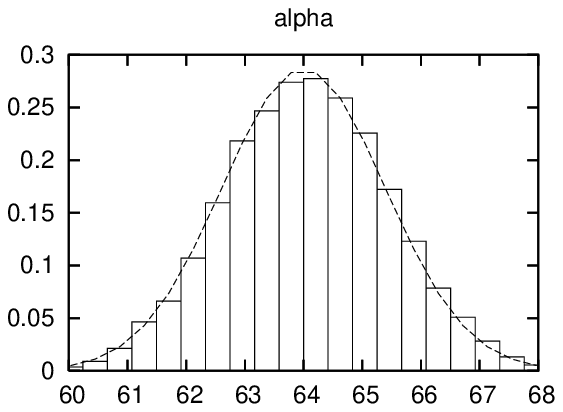} & \includegraphics{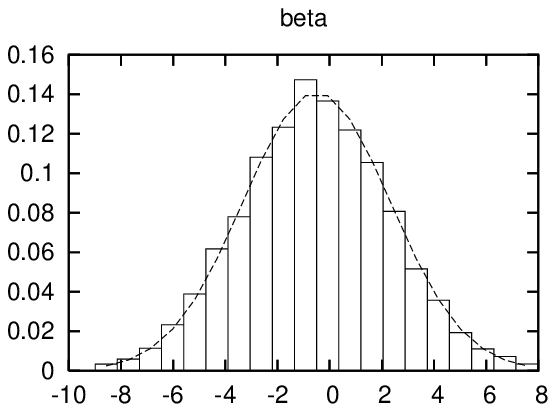}\\
\includegraphics{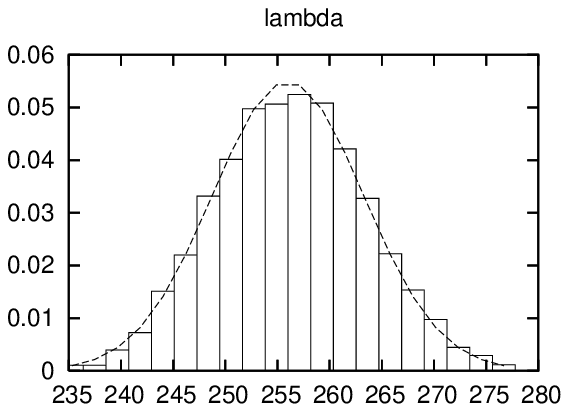} & \includegraphics{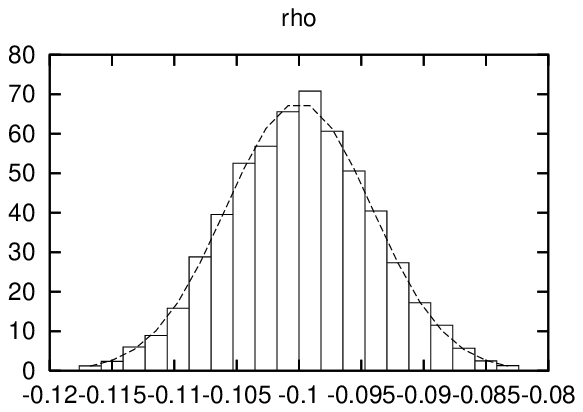}\\
\end{array}
$$
\caption{Empirical and asymptotic distribution of the simple
estimators for the $\Gamma$-OU model. The histograms are produced
from $m=10000$ replications consiting of $n=8000$ observations
each, corresponding to 32 years with 250 daily observations per
year. The true values are $\nu=2.56$, $\alpha=64$, $\lambda=256$,
$\mu=1.2$, $\beta=-0.5$, $\rho=-0.1$. The standard deviations used
for the normal curves are taken from the explicit asymptotic
results, not estimated.}
\end{figure}

Figure~\ref{his} illustrates the empirical and asymptotic
distribution of the simple estimators for the $\Gamma$-OU model.
The histograms are produced from $m=10000$ replications consisting
of $n=8000$ observations each, corresponding to 32 years with 250
daily obervations per year.
We see from the graphs that in our illustration the parameters
$\nu$, $\alpha$, $\lambda$, and $\mu$ can be estimated quite
accurately, in the sense that the usual confidence intervals
yield one or two significant digits at least.
The estimate for $\rho$ is not as accurate and the
accuracy for the estimate for $\beta$ is unsatisfactory.

The bad quality of the estimator for $\beta$ is neither surprising
nor very troublesome. It has little impact on the model. The main
reason for including the parameter $\beta$ in the specification of
BNS models is, for derivatives pricing: A risk-neutral BNS-model
must have $\beta=-1/2$. In most applications working under a
physical probability measure $\beta=0$ can be assumed without much
loss of generality or flexibility.

In ongoing work \cite{HP2006} we compare this asymptotic covariance
with the covariance of the optimal quadratic estimating function.
\section{Further and alternative developments\label{further}}
\subsection{Optimal quadratic estimating functions}
Our choice of estimating functions is natural, but, mathematically
speaking, somewhat arbitrary. In ongoing work \cite{HP2006} we show,
that the {\em optimal quadratic estimating function} based on the
moments of $V_1,X_1,V_1^2,V_1X_1,X_1^2$ can be computed
explicitly, though the corresponding estimator has to be
determined numerically. Our simple estimator can be used as a
starting point for an iterative root-finding procedure.
Consistency and asymptotic normality can be shown using the
general theory as presented in \cite{SO99} along the lines of
the present paper, although the expressions involved are slightly
more complicated.
\subsection{Using more integer or trigonometric
moments for better efficiency} More efficient estimators than
provided by the optimal quadratic estimating function can be
obtained by incorporating further moments. As we have provided
explicit computations for arbitrary integer moments and
conditional moments, our methods can be extended to that
situation. We might even have the number of moments tend to
infinity with the number of observations, and obtain an estimator
that is asymptotically equivalent to the maximum likelihood
estimator, when the latter exists resp.\ can be defined,
see~\ref{remarks}. The reader might object, that very high moments
are not reliable for empirical investigations. BNS-models allow
also explicit computation of the characteristic function and thus
of conditional and unconditional {\em trigonometric moments}
$E[e^{i(\xi_kV_1+\psi_kX_1)}]$ and
$E[e^{i(\xi_kV_1+\psi_kX_1)}|V_0]$ for arbitrary constants $\xi_k$
and $\psi_k$, that could be used instead for constructing
estimating functions. See \cite{ASa2002} for diffusions,
\cite{Sch2005} for \Levy{} type processes, and \cite{Sin2001} for
affine models.
\subsection{Intra-day observations}
Our approach is based on the explicit calculation of conditional
and unconditional moments. Those calculations can be done for
BNS-models on arbitrary time intervals. Hence our analysis is not
restricted to a fixed time grid with the number of observation
intervals tending to infinity, but could be performed also on a
fixed horizon, with the number of intra-day observations
increasing to infinity. The resulting estimators should then be
compared to power-variation methods, cf.~\cite{Tod2006}.
\subsection{Comparison to the generalized method of moments}
We would be interested in a comparison of our results to the
related {\em generalized methods of moments}. For a rigorous
treatment of the latter, a precise specification of the weighting
matrix is required, see \cite{HHY1996} and the references therein.
\subsection{Unobserved volatility and substitutes for volatility}
Finally, perhaps the biggest issue is, that the instantaneous
variance is not observed in discrete time.
In \cite{Lindberg1} it is reported, that the number of trades
is an excellent substitute for statistical purposes.
This is certainly a promising starting point for an empirical
analysis. For a theoretical analysis a joint model for the number
prices and number of trades has to be specified.

Another direction would be to adapt the implied state method
(IS-GMM) as introduced in \cite{Pan} to our martingale estimating
function approach: We replace the unobserved $V_i$ in the
estimating equations by the model-implied variance $V_i(\theta)$
that is obtained from option prices, assuming that the dynamics
are governed by BNS-models both under the physical probability
measure $P_{\theta_0}$ and a risk-neutral measure
$P_{\tilde\theta_0}$. The resulting estimating function will not
be a martingale estimating function any more, and the bias has to
be accounted for in a rigorous analysis. Nevertheless, in view of
the results of \cite{Pan}, we are optimistic, that consistency and
asymptotic normality will hold also here.
\appendix
\section{Explicit moment calculations\label{explicit}}\label{ApA}
This section is about computing explicitly $E[X_1^nV_1^m|V_0=0]$
and $E[X_1^nV_1^m]$.
All moments below will be given in terms of the cumulants of the
stationary distribution, denoted by $K_n$.
We set
\begin{equation}
\zeta=K_1,\qquad
\eta=K_2.
\end{equation}
If the stationary distribution is determined by the two parameters $\zeta$ and
$\eta$ the higher cumulants are obviously functions of $\zeta$ and $\eta$,
but the formulae hold in more general cases.

The calculations exploit the analytical tractability of the BNS-model,
namely conditional Gaussianity of the logarithmic returns $X$ and
the linear structure of the OU-type process~$V$.
From that it follows, and it is well-known, that univariate and
multivariate cumulants can be computed easily.
It remains to transform multivariate cumulants to multivariate moments,
again a topic that is well-understood,
and explicit expressions involve the multivariate Faa di Bruno formula,
multivariate Bell polynomials and integer partitions,
see for example~\cite{McC1987}.

We have chosen to use simple recursions, that are easy to
implement on a computer algebra system, in particular, since the
expressions, though completely explicit and elementary, are rather
lengthy when it comes to evaluating moments of order four for the
asymptotic covariance matrix. For the reader's convenience, we
give the details in this appendix.
\subsection{Preliminaries}
Let us recapitulate the variables and notation from
section~\ref{discrete}, that are required in the following
calculations. We use
\begin{equation}
\gamma=e^{-\lambda\Delta},\qquad
\epsilon=\frac{1-e^{-\lambda\Delta}}{\lambda}.
\end{equation}
We have
\begin{equation}\label{VY-reduce}
V_1=\gamma V_0+U_1,\qquad
Y_1=\epsilon V_0+S_1
\end{equation}
where
\begin{equation}
U_1=\int_0^\Delta e^{-\lambda(\Delta-s)}dZ_{\lambda s},\qquad
S_1=\int_0^\Delta \lambda^{-1}(1-e^{-\lambda(\Delta-s)})dZ_{\lambda s}.
\end{equation}
Note, that we have the simpler formula $S_1=(Z_1-U_1)/\lambda$, but the integral above is sometimes notationally more convenient.
We have
\begin{equation}\label{X-reduce}
X_1=A_1+\sqrt{Y_1}W_1,\qquad
A_1=\mu\Delta+\beta Y_1+\rho Z_1.
\end{equation}
\subsection{Stationary moments}
We use the well-known recursion to compute moments from cumulants
%
%
\begin{equation}
E[V_0^n]=\delta_{n0}+\sum_{i=0}^{n-1}\binom{n-1}{i}K_{i+1}E[V_0^{n-1-i}].
\end{equation}
Alternatively we have $E[V_0^n]=Y_n(K_1,\ldots,K_n)$, where $Y_n(x_1,\ldots,x_n)$
denotes the complete Bell polynomials.
%
%
Explicit non-recursive expressions can be given,
but we do not use them.
\subsection{Trivariate cumulants}
From the key formula for Wiener-type integrals with \Levy{} process integrator,
it follows that the joint cumulants of $(S_1,U_1,Z_1)$ are given by
\begin{equation}
K_{nm\ell}=\lambda\epsilon_{nm}(n+m+\ell)K_{n+m+\ell},
\end{equation}
with
\begin{equation}
\epsilon_{ij}=\left\{\begin{array}{ll}
\displaystyle\lambda^{-i}\left(1+\sum_{k=1}^i\binom{i}{k}(-1)^k\frac{1-\gamma^k}{k\lambda}\right) & j=0\\
\\
\displaystyle\lambda^{-i}\left(\frac{1-\gamma^j}{j\lambda}+\sum_{k=1}^i\binom{i}{k}(-1)^k\frac{1-\gamma^k}{k\lambda}\right) & j>0
\end{array}\right.
\end{equation}
\subsection{Trivariate Moments}
Trivariate moments can be computed recursively from trivariate cumulants
%
%
\begin{equation}
E[S_1^nU_1^mZ_1^\ell]=
\sum_{i=0}^{n-1}\sum_{j=0}^m\sum_{k=0}^{\ell}\binom{n-1}{i}\binom{m}{j}\binom{\ell}{k}
K_{i+1,j,k}E[S_1^{n-1-i}U_1^{m-j}Z_1^{\ell-k}]
\end{equation}
\begin{equation}
E[S_1^nU_1^mZ_1^\ell]=
\sum_{i=0}^{n}\sum_{j=0}^{m-1}\sum_{k=0}^{\ell}\binom{n}{i}\binom{m-1}{j}\binom{\ell}{k}
K_{i,j+1,k}E[S_1^{n-i}U_1^{m-1-j}Z_1^{\ell-k}]
\end{equation}
\begin{equation}
E[S_1^nU_1^mZ_1^\ell]=
\sum_{i=0}^{n}\sum_{j=0}^m\sum_{k=0}^{\ell-1}\binom{n}{i}\binom{m}{j}\binom{\ell-1}{k}
K_{i,j,k+1}E[S_1^{n-i}U_1^{m-j}Z_1^{\ell-1-k}]
\end{equation}
Alternatively, we can express $E[S_1^nU_1^mZ_1^\ell]$ as
trivariate complete Bell polynomials
$Y_{nm\ell}$ evaluated at the trivariate cumulants of $S_1,U_1,Z_1$,
and explicit non-recursive expressions are available,
but not very useful for us.
%
%
\subsection{Some conditional expectations}
Using (\ref{VY-reduce}) gives
\begin{equation}
E[Y_1^nV_1^mZ_1^\ell|V_0=v]=
\sum_{i=0}^n\sum_{j=0}^m\binom{n}{i}\binom{m}{j}\epsilon^i\gamma^jE[S_1^{n-i}U_1^{m-j}Z_1^{\ell}]\cdot v^{i+j}
\end{equation}
Collecting powers of $v$ gives
\begin{equation}
E[Y_1^nV_1^mZ_1^\ell|V_0=v]=
\sum_{k=0}^{n+m}\xi_{nm\ell k}v^k
\end{equation}
with
\begin{equation}
\xi_{nm\ell k}=
\sum_{j=0}^{m\wedge k}\binom{n}{k-j}\binom{m}{j}\epsilon^{k-j}\gamma^jE[S_1^{n-k+j}U_1^{m-j}Z_1^{\ell}]
\end{equation}
Then using (\ref{X-reduce}) and conditioning gives
\begin{equation}
E[A_1^{n}Y_1^{m}V_1^{\ell}|V_0=v]=
\sum_{i=0}^n\sum_{j=0}^{n-i}\binom{n}{i}\binom{n-i}{j}\beta^i\rho^j\mu^{n-i-j}E[Y_1^{m+i}V_1^{\ell}Z_1^{j}|V_0=v]
\end{equation}
Collecting powers of $v$ gives
\begin{equation}
E[A_1^{n}Y_1^{m}V_1^{\ell}|V_0=v]=
\sum_{k=0}^{n+m+\ell}\psi_{nm\ell k}v^k
\end{equation}
with
\begin{equation}
\psi_{nm\ell k}=
\sum_{i=(k-m-l)_+}^{n}\sum_{j=0}^{n-i}\binom{n}{i}\binom{n-i}{j}\beta^i\rho^j\mu^{n-i-j}\xi_{m+i,\ell,j,k}
\end{equation}
Finally using (\ref{X-reduce}) and the Gaussian moments gives
\begin{equation}
E[X_1^nV_1^m|V_0=v]=
\sum_{i=0}^{\lfloor n/2\rfloor}\binom{n}{2i}\frac{(2i)!}{2^ii!}E[A_1^{n-2i}Y_1^{i}V_1^{m}|V_0=v]
\end{equation}
Collecting powers of $v$ gives
\begin{equation}
E[X_1^nV_1^m|V_0=v]=
\sum_{k=0}^{n+m}\phi_{nmk}v^k
\end{equation}
with
\begin{equation}
\phi_{nmk}=
\sum_{i=0}^{(n+m-k)\wedge\lfloor\frac{n}{2}\rfloor}\binom{n}{2i}\frac{(2i)!}{2^ii!}\psi_{n-2i,i,m,k}
\end{equation}
It follows from the calculations above that $\phi_{nmk}$ are polynomials in
$\gamma,\epsilon,\mu,\beta,\rho$.
\subsection{Some unconditional expectations}
The same structure pertains for the unconditional expectations,
\begin{equation}
E[Y_1^nV_1^mZ_1^\ell]=
\sum_{i=0}^n\sum_{j=0}^m\binom{n}{i}\binom{m}{j}\epsilon^i\gamma^jE[S_1^{n-i}U_1^{m-j}Z_1^{\ell}]E[V_0^{i+j}]
\end{equation}
then
\begin{equation}
E[A_1^{n}Y_1^{m}V_1^{\ell}]=
\sum_{i=0}^n\sum_{j=0}^{n-i}\binom{n}{i}\binom{n-i}{j}\beta^i\rho^j\mu^{n-i-j}E[Y_1^{m+i}V_1^{\ell}Z_1^{j}]
\end{equation}
and finally
\begin{equation}
E[X_1^nV_1^m]=
\sum_{i=0}^{\lfloor n/2\rfloor}\binom{n}{2i}\frac{(2i)!}{2^ii!}E[A_1^{n-2i}Y_1^{i}V_1^{m}].
\end{equation}

\section{The simple multivariate martingale central limit theorem\label{multivariate}}
The following simple version of a multivariate martingale central
limit theorem is certainly well-known or obvious for experts,
some references are~\cite{CriPra,KueSoe,vZ}.

However, when looking for references, we found statements that do
not exactly apply, or that are much more general (continuous time,
random normalizations,\ldots). It turned out that the elementary
proof below is shorter, than an attempt to verify the assumptions
and deduce the result from a more 'advanced' theorem. Yet, any
{\em concrete and precise} hint for an appropriate reference would
be most welcome to the authors.
\begin{theorem}Suppose $\dis (X_{n,k})$ is a martingale difference
array such that for every $\epsilon>0$
\begin{equation}\label{uvjet1}\sum_{k=1}^n
E\big[\|X_{n,k}\|^2\mathbb{1}_{\{\|X_{n,k}\|>\epsilon\}}|\lijepof_{k-1}\big]\stackrel{P}{\longrightarrow}
0
\end{equation} and
\begin{equation}\label{uvjet2}\sum_{k=1}^n\big[X_{n,k}X_{n,k}^\top|\lijepof_{k-1}\big]
\stackrel{P}{\longrightarrow}\Upsilon
\end{equation}as $n\to\infty.$
Then
\begin{equation}\sum_{k=1}^n
X_{n,k}\inlawto N(0,\Upsilon).
\end{equation}\hfill
\em \dz We will use the Cramer-Wold device. For $\dis \beta\in
\mathbb{R}^d,\ \beta\neq 0,$ let us define a random variable
\begin{equation}Y_{n,k}=\beta^\top X_{n,k}.
\end{equation}Then we have
\begin{eqnarray}\label{u1}\sum_{k=1}^n
E\big[Y_{n,k}^2|\lijepof_{k-1}\big]&=&\sum_{k=1}^n
E\big[(\beta^\top X_{n,k})(X_{n,k}^\top\beta)|\lijepof_{k-1}|\big]\nonumber\\
&=&
\beta^\top\sum_{k=1}^nE\big[X_{n,k}X_{n,k}^\top|\lijepof_{k-1}]\beta.
\end{eqnarray}
From Assumption (\ref{uvjet2}) it follows that the expression
(\ref{u1}) converges to $\beta^\top\Upsilon\beta$ and thus
\begin{equation}\label{MLT1}\sum_{k=1}^n
E\big[Y_{n,k}^2|\lijepof_{k-1}\big]\to\beta^\top\Upsilon\beta
\end{equation}as $n\to\infty.$
Furthermore, it holds
\begin{equation}\big|\beta^\top X_{n,k}|\leq
\|\beta\|\cdot\|X_{n,k}\|.
\end{equation}
Thus, for an arbitrary $\epsilon>0$ we have
\begin{eqnarray}\label{MLT2}0&\leq&
\sum_{k=1}^nE\big[Y_{n,k}^2\mathbb{1}_{\{|Y_{n,k}|>\epsilon\}}|\lijepof_{k-1}\big]\nonumber\\
&=&\sum_{k=1}^nE\big[(\beta^\top
X_{n,k})^2\mathbb{1}_{\{|\beta^\top
X_{n,k}|>\epsilon\}}|\lijepof_{k-1}\big]\nonumber\\
&\leq&
\|\beta\|^2\sum_{k=1}^nE\big[\|X_{n,k}\|^2\mathbb{1}_{\{|\beta^\top
X_{n,k}|>\epsilon\}}|\lijepof_{k-1}\big].
\end{eqnarray}
Since for $\beta\neq 0$ the condition $|\beta^\top
X_{n,k}|>\epsilon$ implies
\begin{equation}\|X_{n,k}\|\geq\|\beta\|^{-1}|\beta^\top
X_{n,k}|>\|\beta\|^{-1}\epsilon,
\end{equation}it follows that
\begin{equation}\mathbb{1}_{\{|\beta^\top
X_{n,k}|>\epsilon\}}\leq\mathbb{1}_{\{\|X_{n,k}\|>\|\beta\|^{-1}\epsilon\}}
\end{equation}and thus using the Assumption (\ref{uvjet1}), from
(\ref{MLT2}) it follows that
\begin{equation}\sum_{k=1}^nE\big[Y_{n,k}^2\mathbb{1}_{\{|Y_{n,k}|>\epsilon\}}|\lijepof_{k-1}\big]
\stackrel{P}{\longrightarrow} 0
\end{equation}as $n\to\infty.$
Now, the statement follows from the univariate martingale central
limit theorem from \cite{HH80}.
\qed
\end{theorem}
\begin{lemma}The conditional Lyapounov condition implies the
conditional Lindeberg condition, namely, if
\begin{equation}\label{LJ}\sum_{k=1}^nE\big[\|X_{n,k}\|^4|\lijepof_{k-1}\big]\longrightarrow
0 \end{equation} then
\begin{equation}\sum_{k=1}^n
E\big[\|X_{n,k}\|^2\mathbb{1}_{\{\|X_{n,k}\|>\epsilon\}}|\lijepof_{k-1}\big]\longrightarrow
0
\end{equation}as $n\to\infty.$

\noindent\proof  For every $\epsilon>0$ we have
\begin{eqnarray}\sum_{k=1}^nE\big[\|X_{n,k}\|^4|\lijepof_{k-1}\big]&=&
\sum_{k=1}^nE\big[\|X_{n,k}\|^4\mathbb{1}_{\{\|X_{n,k}\|>\epsilon\}}|\lijepof_{k-1}\big]+
\sum_{k=1}^nE\big[\|X_{n,k}\|^4\mathbb{1}_{\{\|X_{n,k}\|\leq\epsilon\}}|\lijepof_{k-1}\big]\nonumber\\
&\geq&\epsilon^2\sum_{k=1}^nE\big[\|X_{n,k}\|^2\mathbb{1}_{\{\|X_{n,k}\|>\epsilon\}}|\lijepof_{k-1}\big],
\end{eqnarray}since
$$\sum_{k=1}^nE\big[\|X_{n,k}\|^4\mathbb{1}_{\{\|X_{n,k}\|\leq\epsilon\}}|\lijepof_{k-1}\big]\geq 0.$$
\end{lemma}From Assumption (\ref{LJ}) the statement
follows.
\qed
\bibliography{siest1}
\end{document}